\documentclass[pre]{revtex4}
\usepackage{amsmath}
\begin{document}

\title{Validity of the linear marginal stability principle for monotonic fronts of
the extended Fisher-Kolmogorov equation}
\author{R. D. Benguria}
\affiliation{
 Facultad de F\'\i sica\\
    Pontificia Universidad Cat\'olica de Chile\\
           Casilla 306, Santiago 22, Chile}
\author{ M. C. Depassier}
\affiliation{
 Facultad de F\'\i sica\\
    Pontificia Universidad Cat\'olica de Chile\\
           Casilla 306, Santiago 22, Chile}

\date{\today}

\begin{abstract}
The extended Fisher Kolmogorov equation  $u_t = u_{xx} - \gamma
u_{xxxx} + f(u)$ with arbitrary positive $f(u)$, satisfying $f(0) =
f(1) =0$, has monotonic traveling fronts for $\gamma < 1/12$. We
find a simple lower bound on the speed of the fronts which allows
to assess the validity  of linear  marginal stability.
\end{abstract}

\pacs{  05.45-a, 52.35.Mw,  02.60.Lj}
\keywords{
traveling waves, extended Fisher-Kolmogorov equation, fronts}

\maketitle

\section{Introduction}

The extended Fisher-Kolmogorov equation (EFK),
\begin{equation}
u_t = u_{xx} - \gamma u_{xxxx} + f(u),
\label{EFK}
\end{equation}
with $f(u) = u - u^3$ arises in in the description of different
systems. It appears for example in the  study of phase transitions near a
Lifshitz point \cite{HLS,Z}. It has been derived as an amplitude
equation at the onset of instabilities near certain degenerate
points \cite{Rott-Doelman}. It has also been proposed as a model for
the onset of spatiotemporal chaos in bistable systems \cite{CER}
and as a natural extension to the reaction diffusion equation
($\gamma=0$) on which to study the dynamics of front propagation
and pattern formation \cite{Dee-vS,vanSaarloosa} etc.  Its steady
version with different functions $f(u)$,
$$
u_{xxxx} + q u_{xx} + f(u) =0,
$$
is of interest in different fields and much work has been devoted
to it. A very complete account of its solutions can be found in
\cite{Peletier-Troy2001}.

 For $\gamma < 1/12$ numerical results
indicate that sufficiently localized conditions evolve into a
uniform translating front joining the stable point u=1 to the
unstable u=0 point \cite{vanSaarloosb}. Similarly to what is found in the reaction diffusion equation, 
 for the Fisher case \cite{Fisher,KPP}
$f(u) = u - u^2$ and for $f(u) = u - u^3$ the front propagates
with the linear speed which now is \cite{vanSaarloosa}
\begin{equation}
c_L = \frac{2}{\sqrt{54 \gamma}} [1 + 36 \gamma -
(1-12\gamma)^{3/2}]^{1/2}, 
\label{linearspeed}
\end{equation}
obtained from linear analysis  near $u=0$. If $\gamma > 1/12$
monotonic fronts do not exist, and a pattern may appear.

Numerical results of the integrations of the EFK equation with
arbitrary $f(u)$ show that, as it occurs in the reaction diffusion
equation, transition from linear to nonlinear marginal stability
will occur as parameters in $f(u)$ are varied \cite{vanSaarloosb}.

In recent work a sufficient criterion 
$f(u)$ for the validity of the linear speed selection mechanism
analogous to the KPP \cite{KPP} and Aronson-Weinberger \cite{AW78}  criteria  for
the reaction diffusion equation has been established \cite{Ebert2004}. As for the reaction diffusion equation, this criterion gives sufficient but not necessary conditions on the validity of the linear speed selection mechanism. 
 Numerical results indicate that for small $\gamma$ fronts of the EFK equation have similar properties to fronts of
 the reaction diffusion equation. Rigorous existence results of fronts of the EFK equation have been given for general
 functions $f(u)$\cite{Bouwe2001}. For $\gamma \rightarrow 0$ it was proved that there is a minimal speed $c^*$ for the
 existence of monotonic fronts, and that the fronts are stable.
 For $\gamma = \epsilon^2$ the minimal speed is given by $c^* \ge  2 -  \epsilon^2 +.....$\cite{Rott-Wayne2001}.

The purpose of the present work is to establish a simple lower bound on the speed $c^*$ for which monotonic fronts exist.
 This enables to test whether for a given function $f(u)$ the minimal speed is the linear value $c_L$ obtained from the
 linear analysis at the edge of the front. The bound given in this work is not sharp,
 but  the derivation suggests that it is  possible to obtain a variational formulation for the minimal speed
  analogous to that given for the reaction-diffusion equation \cite{BD96a,BD96c}.
Future work will address this aspect.

\section{Monotonic fronts}

A traveling monotonic front $u= q(x - c t)$ joining the stable state $u=1$ to $u=0$
satisfies the ordinary differential equation
$$
q_{zz} +c q_z - \gamma q_{zzzz} + f(q) = 0,
$$
with $\qquad 
\lim_{z\rightarrow-\infty} q = 1, \qquad \lim_{z\rightarrow\infty} q = 0, \qquad q_z < 0,
$

where $z = x-c t$ and subscripts denote derivatives.

Following the usual procedure, since the front is monotonic, we
may use phase space variables and define $p(q) = - dq/dz$, where
the minus sign is introduced to have $p>0$. A simple calculation
shows that monotonic fronts obey
\begin{equation}
\gamma p \frac{d }{d q}\left[ p \frac{d }{d q}\left(p \frac{d p}{d q}\right)\right] - p \frac{d p}{d q} + c p - f(q) = 0
\label{main}
\end{equation}
with
$$
p(0) = p(1) = 0, \quad \mbox{and} \quad p>0.
$$
Next we obtain the upper bound. Let $g(q)$ be an arbitrary positive decreasing function. Multiplying Eq. (\ref{main}) by $g/p$ and integrating with respect to $q$ between 0 and 1 we obtain the identity
\begin{equation}
c \int_0^1 g(q) d q = \int_0^1 \frac{g f}{p} d q + \int_0^1 p\, h\, d q + \gamma \int_0^1 \left( \frac{1}{3} g''' p^3 + h\, p\, p'^2 \right) d q,
\label{identidad}
\end{equation}
where primes denote derivatives with respect to $q$ and where we
have defined $h(q) = -g'(q) > 0$. In obtaining this expression
several integrations by parts were performed. Surface terms vanish
due to the boundary conditions on $p$. Furthermore we assume that
the function $g$ does not diverge in a manner that prevents the
vanishing of surface terms.

Consider now the functional
\begin{equation}
S_g(p) = \int_0^1 \frac{g f}{p} d q + \int_0^1 p h d q + \gamma \int_0^1 \left( \frac{1}{3} g''' p^3 + h p p'^2 \right) d q.
\label{accion}
\end{equation}
It can be shown (details will be given elsewhere)
 that for $g \in C^3([0,1]), g'<0, g'''>0$, this functional  has a unique minimizer which we call $\hat p$. 
 Therefore
$$
S_g(p) \ge \min_p S_g(p) = S_g(\hat p).
$$
This implies in Eq. (\ref{identidad}) that
\begin{equation}
c \int_0^1 g(q) d q \ge S_g(\hat p).
\label{ineq1}
\end{equation}
The minimizing $p$, $\hat p$ can be obtained by solving the Euler-
Lagrange equation for $S_g(p)$,
$$
\frac{d}{d q} \left(\frac{ \partial L}{\partial p'}\right) -
\frac{
\partial L}{\partial p} = 0.
$$
Recalling that the arbitrary function $g$ is a function of $q$ we
obtain
\begin{equation}
2 \gamma \frac{d}{d q} [ h \hat p \hat p'] + \frac{ g f}{\hat p^2}
+ g' + \gamma (g'\hat p'^2 - g''' \hat p^2) = 0. \label{Euler}
\end{equation}
To obtain the minimizing $p$ for each function $g(q)$ we should solve this
equation. This is not an easy task since $g(q)$ is an arbitrary
unspecified function. However, it follows from this equation, multiplying
by $p(q)$ and integrating in $q$ that
$$
\int_0^1 \left(\frac{g f}{\hat p} - h\, p - \gamma g''' p^3 - 3 \gamma
h p'^2 p \right)d q  = 0.
$$
Using this result we find that $S_g(\hat p)$ can be written as
$$
S_g(\hat p) = \frac{4}{3} \int_0^1 \frac{ f g}{\hat p} + \frac{2}{3}
\int_0^1 \hat p\,  h \, d q.
$$
Inequality  (\ref{ineq1}) is then
$$
c \int_0^1 g(q) d q \ge \frac{4}{3} \int_0^1 \frac{ f(q)
g(q)}{\hat p(q)} d q  + \frac{2}{3} \int_0^1 \hat p(q) h(q) \, dq.
$$
Finally, since $f>0, g>0$ and $h = -g' > 0$ we use the inequality
$a^2 + b^2 \ge 2 a b$ in the expression above to obtain our main
result,
\begin{equation}
c \ge \frac{ 4 \sqrt{2}}{3} \frac{ \int_0^1 \sqrt{ f g h} d q}{
\int_0^1 g dq}. \label{final}
\end{equation}
Here we recall that $g$ is an arbitrary decreasing monotonic
function.

Notice that this expression is similar in form to the bound
obtained previously for the speed of fronts of the reaction
diffusion equation. In that case we proved that the bound is sharp
and that it follows from a variational principle. In the present
case  this bound does not saturate, however a variational
principle will follow from (\ref{ineq1}) if we succeed in proving
that there is a certain function $g(q)$ for which $\hat p$ is the
solution of the differential equation (\ref{main}). This point
will be addressed in future work.

\section{Conclusion}

A lower bound on the speed of monotonic fronts of the EFK equation has been obtained. 
This bound allows to determine the range of validity of the linear speed selection mechanism.
We conjecture that there is a variational principle for the minimal speed of the fronts 
from which its exact value could be calculated.

\section*{Acknowledgments}

R. D. Benguria and M. C. Depassier
acknowledge partial support from
 Fondecyt under grants 1020844 and 1020851.

\end{document}